# Leakage Mitigation and Internal Delay Compensation in FMCW Radar For Small Drone Detection

Junhyeong Park, Seungwoon Park, and Seong-Ook Park, *Senior Member, IEEE*

*Abstract*—One of the notorious problems of frequency modulated continuous-wave (FMCW) radar is leakage between the transmitter and the receiver. The phase noise of the leakage is expressed as a skirt around the leakage signal on power spectrum. It causes the deterioration of the dynamic range, especially, in the near-distance region. Therefore, although FMCW radar has an advantage over pulse radar in terms of near-distance target detection due to its way of operation, the advantage of FMCW radar can be lost because of the leakage. Another problem of FMCW radar is internal delay in the radar system. It leads to the decrease of the maximum detectable range. In this paper, a novel down-conversion technique which resolves these problems is proposed. Detailed theory and procedures to implement the proposed technique are explained. Then, performances of it are verified with the experiment results. The proposed technique can be implemented through frequency planning and digital signal processing without additional parts. The results show that the proposed technique lowers the noise floor about 7.0 dB in the near-distance region and 2.1 dB even in the far-distance region. Also, the results demonstrate the proposed technique recover the reduced maximum detectable range by compensating the internal delay.

## I. INTRODUCTION

MANY studies on frequency modulated continuous-wave (FMCW) radar have been carried out for decades. Nowadays, FMCW radar is widely used for various applications. Because FMCW radar can measure the distance, it is used in level meter and altimeter [1]-[3]. Not only can it measure the distance, but also analyze Doppler frequency. Thus it is also applied in moving target detection radar, vital sign detection radar, meteorological radar, and synthetic aperture radar (SAR) [4]-[8]. In comparison with pulse radar, FMCW radar has advantages in cost, peak power, and minimum detectable range. Especially, the minimum detectable range is inevitable in pulse radar since the receiver is turned off while the pulse is being emitted [9]. On the other hand, FMCW radar continuously receives the electromagnetic wave. Therefore, the operation mechanism of FMCW gives a benefit in near-distance target detection.

Although FMCW radar has many advantages, there is a notorious problem called leakage. In monostatic FMCW radar, the leakage occurs in a circulator due to its inefficient isolation capability, and the mismatch in antenna also leads to the leakage [11]-[13]. In case of bistatic FMCW radar, the transmitted signal is leaked into the receiver through spatial paths between antennas, and objects near antennas can also be the cause of the leakage [10]. The leakage produces severe problems. Because its power is typically much higher than that of returned signals, low-noise amplifier (LNA) can be saturated [10], [11]. Besides, the dynamic range of FMCW radar is limited by the phase noise of the leakage [11]. On the power spectrum after deramping and down-conversion process, the phase noise of the leakage is expressed as a skirt around the leakage signal. It raises overall noise floor and the noise floor near the leakage signal is significantly raised. Thus, the phase noise of the leakage generally decreases signal-to-noise ratio (SNR), and SNR at the near-distance region is critically damaged. For example, if there is a small target in the near-distance, the beat signal of the small target can be buried under the phase noise skirt. Therefore, even though FMCW radar has the benefit in the near-distance target detection, if the phase noise problem of the leakage is not resolved, the benefit can be diminished.

Another problem in FMCW radar is internal delay in the radar system. The desired distance information from FMCW radar is the distance between the antenna and the target. This distance can be obtained by using relations between time-delay, beat frequency, and distance. However, there are many parts in FMCW radar system, and these produce inevitable delays. Therefore, in practice, the internal delay in FMCW radar system is fundamentally added, and it means that undesired additional beat frequency occurs. This causes the reduction of the maximum detectable range. As the internal delay increases, the problem gets worse. If quite long internal delay is necessary in FMCW radar [18], [19], the critical reduction of the

This paragraph of the first footnote will contain the date on which you submitted your paper for review. It will also contain support information, including sponsor and financial support acknowledgment. This work was supported by a project of UAV Safety Technology Research called "Flight Safety Regulation Development and Integrated Operation Demonstration for Civil RPAS" funded by the Ministry of Land, Infrastructure and Transport (MOLIT), Republic of Korea Government under Grant 18ATRP-C000000-00.

The authors are with the School of Electrical Engineering, Korea Advanced Institute of Science and Technology (KAIST), Daejeon 34141, Republic of Korea.

maximum detectable range occurs, so its compensation is important.

There have been many studies to resolve the leakage problem in FMCW radar. Beasley *et al.* proposed adding a closed loop leakage canceller for the monostatic radar [12]. The closed loop leakage canceller adaptively generates an error vector which includes amplitude and phase information of the leakage. Lin *et al.* implemented the scheme of Beasley *et al.* in digital signal processing (DSP) approach, and improved it by up-converting the error signal to a preselected reference frequency to overcome the DC-offset problem [11], [13]. Melzer *et al.* studied reflections caused by fixed objects in front of the antennas [14]-[17]. These reflections were introduced as short-range (SR) leakage. Based on the correlation statistics of decorrelated phase noise, Melzer *et al.* selected delay time and implemented it by adding an artificial on-chip target in radio frequency (RF) stage. Then, after mixing expected parameters in digital intermediate frequency (IF) stage, they generated a sampled IF signal which is similar to the sampled IF signal of the SR leakage. Melzer *et al.* tried to cancel the SR leakage by subtracting the two IF signals from each other. Suh *et al.* and Shin *et al.* attempted to reduce the leakage by putting the transmitter (TX) apart from the receiver (RX) [18], [19]. To increase the distance between the TX and the RX, Suh *et al.* used 75 Ω coaxial cables whose lengths are 10 m for TX and 30 m for RX. In case of Shin *et al.*, they used 106.2 m fiber-optic cables that have low-loss characteristic.

Among the problems of the leakage that we mentioned above, we have focused on the phase noise of the leakage through dominant spatial path between antennas and studied how to mitigate it. Also, we have considered how to compensate the internal delay. In the recent conference paper, we proposed a new down-conversion concept to mitigate the phase noise of the leakage in FMCW radar [20]. However, more theoretical analyses to verify the proposed concept were required, and there were only simple simulation results, which are not enough to demonstrate the actual effectiveness of the proposed concept. Due to the short of practical implementation procedures, the proposed concept could not be applied well in practice. Also, the problem of the internal delay was not addressed. Therefore, in this paper, we provides mathematical, conceptual, and experimental analyses of the proposed down-conversion concept in detail. Also, we introduce all detailed procedures to implement the proposed down-conversion technique in practice. In the problematic situations that are the leakage and internal delay problems, the performance of the proposed technique is demonstrated by the experiment results.

The proposed down-conversion technique extracts the frequency and the constant phase information of the beat signal of the leakage in digital IF domain. After generating a digital numerically controlled oscillator (NCO) which has the extracted frequency and constant phase values of the beat signal of the leakage, the received signal is finally down-converted with the digital NCO. By doing this, the frequency and the constant phase in the beat signal of the leakage are removed, so the term of the major cause of the phase noise skirt is removed, and the phase noise of the leakage is concentrated on the stationary point of the cosine function. Therefore, the aforementioned problem on the power spectrum due to the phase noise of the leakage can be significantly attenuated. Also, because the beat frequency of the leakage is mostly due to the internal delay in FMCW radar, the internal delay can be compensated by removing the beat frequency of the leakage. Unlike the previous techniques, the proposed technique can be implemented through frequency planning and DSP without additional parts.

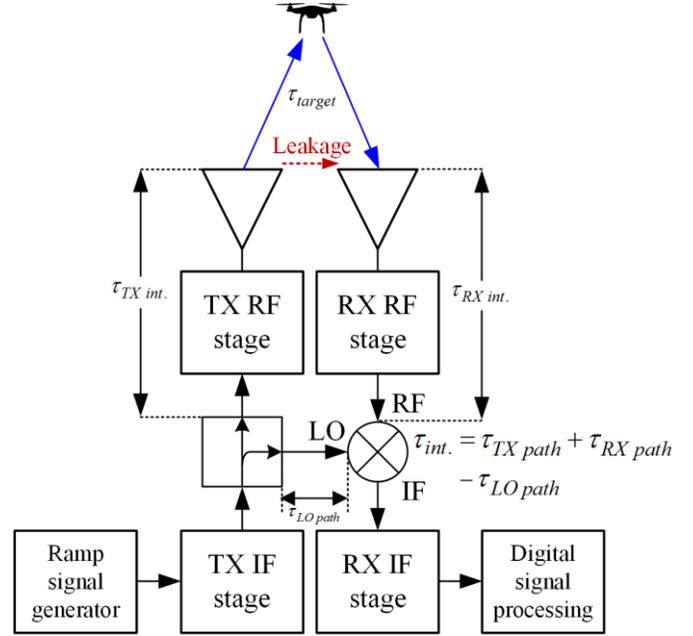

Fig. 1. Block diagram of a common FMCW radar.

For the verification, the results of three experiments are shown. First, I show how much the noise floor is decreased in a situation that only the leakage exists without any target. Second, with a small target which is a drone, DJI Spark, we demonstrate the increased SNR. Lastly, we show that the proposed technique recover the reduced maximum detectable range by compensating the internal delay.

In Section II, principles of common FMCW radar and the proposed down-conversion technique are explained. Detailed procedures of the proposed technique are introduced in Section III. Section IV presents an FMCW radar that we used for the experiments. Then, the three experiments for the verification of the proposed technique are explained In Section V. Results and discussion of the experiments are presented in Section VI, followed by conclusion in Section VII.

## II. PRINCIPLES OF TECHNIQUES

### A. Common Procedures in FMCW Radar

Fig. 1 shows the block diagram of FMCW radar. FMCW radar usually uses a linear frequency modulated (LFM) signal, which is also called as ramp signal or chirp signal. This LFM signal is split into two paths. One passes TX RF stage which may include cables, mixers, filters, isolators, and power amplifiers (PAs), then it is radiated through TX antenna. The other path is mixed with the received LFM signals. We call this LFM signal which is on the other path as reference LFM signal. With this mixing process which is called deramping, we can extract the beat signals which include distance, and even Doppler information of targets. Therefore, the deramping process is one of the most important procedures in FMCW radar. Fig. 2 shows the deramping process.

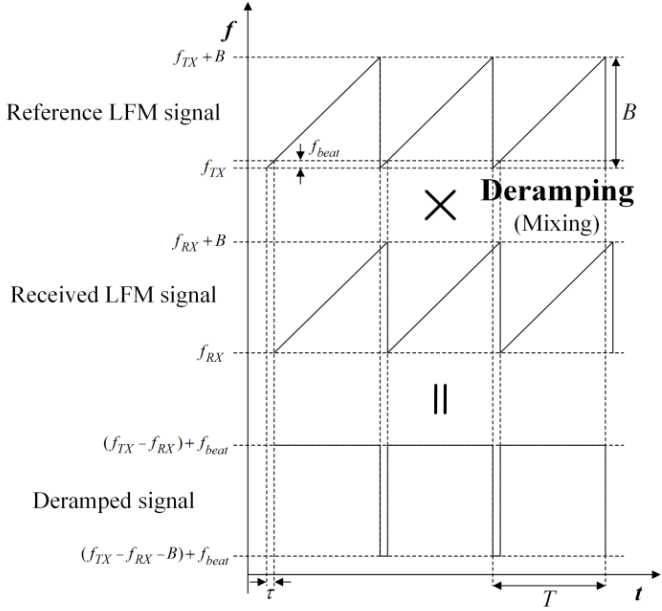

Fig. 2. Deramping process.

In Fig. 1, the LFM signal at the LO port as a reference LFM signal, $s(t)$, can be defined as follows:

$$s(t) = A \cos\left(2\pi f_{TX} t + \pi \frac{B}{T} t^2 + \theta_A + \varphi_A(t)\right) \tag{1}$$

for $0 < t < T$, where $A$ and $f_{TX}$ are the amplitude and the start frequency of the reference LFM signal, $B$ and $T$ are sweep bandwidth and sweep period, $\theta_A$ is constant phase, and $\varphi_A(t)$ is phase noise. The delay from the splitter to the mixer for deramping, i.e., $\tau_{LO\ path}$, is considered together with the other delays at the RF port.

Passing the TX RF stage, the other split LFM signal is delayed by $\tau_{TX\ path,}$. After being transmitted from the TX antenna, the leaked LFM signal directly enters the RX antenna through spatial paths. Then, the signal reflected by the target follows. After being received through the RX antenna, these signals pass the RX RF stage which may include low noise amplifiers (LNAs), filters, isolators, mixers, and cables, so another internal delay, $\tau_{RX\ path}$, is added. For the convenience of derivation, we assume that only a dominant spatial path for the leakage and a dominant single scattering of a fixed target in far-field exist, and time delay due to the spatial path is negligible in the quasi-monostatic radar. Therefore, the received signals at the RF port, $r(t)$, can be expressed as follows:

$$r(t) = \underbrace{B \cos\left(2\pi f_{RX}(t - \tau_{int.}) + \pi \frac{B}{T}(t - \tau_{int.})^2 + \theta_B + \varphi_B(t)\right)}_{Leakage}$$

$$+ \underbrace{C \cos\left(2\pi f_{RX}(t - \tau_{int.} - \tau_{target}) + \pi \frac{B}{T}(t - \tau_{int.} - \tau_{target})^2 + \theta_C + \varphi_C(t)\right)}_{Target}, \tag{2}$$

where $B$ & $C$, $\theta_B$ & $\theta_C$, and $\varphi_B(t)$ & $\varphi_C(t)$ are amplitudes, constant phases, phase noises of the leakage and the target LFM signals at the RF port respectively, $f_{RX}$ is start frequency at the RF port, $\tau_{int.}$ is total internal delay, $\tau_{int.} = \tau_{TX\ path} + \tau_{RX\ path} - \tau_{LO\ path}$, and $\tau_{target}$ is round-trip delay to the target. $B = AG_{TX\ RF}G_{TX\ ant.}PL_{leakge}G_{RX\ ant.}G_{RX\ RF}$ and $C = AG_{TX\ RF}G_{TX\ ant.}PL_{target}\sigma G_{RX\ ant.}G_{RX\ RF}$, where $G_{TX\ RF}$ and $G_{RX\ RF}$ are the total gain of the TX RF stage and the RX RF stage, $G_{TX\ ant.}$ and $G_{RX\ ant.}$ are the gain of the TX and the RX antennas, $PL_{leakge}$ is the near-field path loss of the spatial path for the leakage, $PL_{target}$ is far-field path loss of the round-trip distance for the target, and $\sigma$ is the range cross section of the target.

The phase noises of other local oscillator (LO) signals in RF stage, $\varphi_{TX\ RF\ LO}(t)$, $\varphi_{RX\ RF\ LO}(t)$ can exist when the radar designer wants up-conversion and down-conversion in the TX RF stage and RX RF stage. Therefore, $\varphi_B(t)$ and $\varphi_C(t)$ can be $\varphi_B(t) \approx \varphi_A(t - \tau_{int.}) + \varphi_{TX\ RF\ LO}(t - \tau_{int.}) - \varphi_{RX\ RF\ LO}(t - \tau_{RX\ path})$ and $\varphi_C(t) \approx \varphi_A(t - \tau_{int.} - \tau_{target}) + \varphi_{TX\ RF\ LO}(t - \tau_{int.} - \tau_{target}) - \varphi_{RF\ LO}(t - \tau_{RX\ path})$.

Finally, after the deramping, if we consider only desired terms which are IF beat signals, then deramped signal at the IF port, $x(t)$, can be written as

$$x(t) = x_{IF\ leakage}(t) + x_{IF\ target}(t)$$

$$= \frac{AB}{2}\cos\left(2\pi(\underbrace{\underbrace{f_{TX} - f_{RX}}_{f_{IF\ carrier}} + \underbrace{\frac{B}{T}\tau_{int.}}_{f_{beat,leakage}})t}_{f_{IF\ beat,leakage}}\right.$$

$$\left. + \underbrace{\theta_A + 2\pi f_{RX}\tau_{int.} - \pi \frac{B}{T}\tau_{int.}^2 - \theta_B}_{\theta_{IF\ leakage}} + \underbrace{\varphi_A(t) - \varphi_B(t)}_{\varphi_{IF\ leakage}(t)}\right)$$

$$+ \frac{AC}{2}\cos\left(2\pi(\underbrace{\underbrace{f_{TX} - f_{RX}}_{f_{IF\ carrier}} + \underbrace{\frac{B}{T}\tau_{int.}}_{f_{beat,leakage}})t}_{f_{IF\ beat,leakage}} + \underbrace{\frac{B}{T}\tau_{target}}_{f_{beat,target}}\right)t$$

$$\underbrace{+\theta_A + 2\pi f_{RX}(\tau_{\text{int.}} + \tau_{target}) - \pi\frac{B}{T}(\tau_{\text{int.}} + \tau_{target})^2 - \theta_C}_{\theta_{IF\,target}}$$

$$\left.\underbrace{+\varphi_A(t) - \varphi_C(t)}_{\varphi_{IF\,target}(t)}\right). \quad (3)$$

If $f_{TX} = f_{RX}$, there is no further carrier frequency, but radar designers often make $f_{TX} \neq f_{RX}$ to take advantages in IF stage [18], [19]. In (3) and Fig. 2, we assume $f_{TX} > f_{RX}$, so the beat frequency is added to the IF carrier frequency, $f_{IF\,carrier}$. If $f_{TX} < f_{RX}$, the beat frequency is subtracted from an IF carrier frequency, $|f_{TX} - f_{RX}|$. We can rewrite (3) as follows:

$$x(t) = x_{leakage}(t) + x_{target}(t)$$

$$= \frac{AB}{2}\cos\big(2\pi(f_{IF\,carrier} + f_{beat,leakage})t$$

$$+ \theta_{IF\,leakage} + \varphi_{IF\,leakage}(t)\big)$$

$$+ \frac{AC}{2}\cos\big(2\pi(f_{IF\,carrier} + f_{beat,leakage} + f_{beat,target})t$$

$$+ \theta_{IF\,target} + \varphi_{IF\,target}(t)\big). \quad (4)$$

After passing some blocks in the IF stage, a down-conversion to get rid of the IF carrier frequency and extract beat signals is carried out. Because the IF carrier frequency, $f_{IF\,carrier} = f_{TX} - f_{RX}$, is decided when the radar designers conduct the frequency planning of a radar, the designers already know what the IF carrier frequency is. Thus, $f_{IF\,carrier}$ is chosen as the frequency of the last LO, that is

$$LO_{common}(t) = D\cos\big(2\pi f_{IF\,carrier}t + \theta_{LO} + \varphi_{LO}(t)\big), \quad (5)$$

where $D$, $\theta_{LO}$, and $\varphi_{LO}(t)$ are the amplitude, the constant phase, and the phase noise of the last LO respectively. If (4) is mixed with (5) and passes a low pass filter, then the final signal, $y(t)$, in common FMCW radar can be represented as follows:

$$y(t) = y_{leakage}(t) + y_{target}(t)$$

$$= \underbrace{Z_{leakage}\cos\big(2\pi f_{beat,leakage}t + \theta_{leakage} + \varphi_{leakage}(t)\big)}_{Leakage}$$

$$+ \underbrace{Z_{target}\cos\big(2\pi(f_{beat,leakage} + f_{beat,target})t + \theta_{target} + \varphi_{target}(t)\big)}_{Target}.$$

$$(6)$$

In (6), $Z_{leakage} = ABD/4$, $Z_{target} = ACD/4$, $\theta_{leakage} = \theta_{IF\,leakage} - \theta_{LO}$, $\theta_{target} = \theta_{IF\,target} - \theta_{LO}$, $\varphi_{leakage}(t) = \varphi_{IF\,leakage}(t) - \varphi_{LO}(t)$, and $\varphi_{target}(t) = \varphi_{IF\,target}(t) - \varphi_{LO}(t)$. Then, according to the cosine sum identity, $y_{leakage}(t)$ in (6) can be analyzed as follows:

$$y_{leakage}(t) = Z_{leakage}\cos\big(2\pi f_{beat,leakage}t + \theta_{leakage}\big)\cos\big(\varphi_{leakage}(t)\big)$$

$$- Z_{leakage}\sin\big(2\pi f_{beat,leakage}t + \theta_{leakage}\big)\sin\big(\varphi_{leakage}(t)\big). \quad (7)$$

Generally, since the phase noise is much smaller than 1 radian [14], [20], [26], (7) can be approximated as follows:

$$y_{leakage}(t) \approx Z_{leakage}\cos\big(2\pi f_{beat,leakage}t + \theta_{leakage}\big)$$

$$- \underbrace{Z_{leakage}\varphi_{leakage}(t)\sin\big(2\pi f_{beat,leakage}t + \theta_{leakage}\big)}_{Major\,cause\,of\,the\,phase\,noise\,skirt} \quad (8)$$

As expressed in (8), the phase noise of the leakage signal is up-converted to $f_{beat,leakage}$, and it manifests itself as voltage or current noise. Also, $Z_{leakage}$ is generally much bigger than $Z_{target}$ [10], [11]. These properties create the skirt around the strong leakage signal on the power spectrum. It generally raises the noise floor, and the noise floor around the leakage signal is significantly increased. Besides, since all beat signals contain the beat frequency derived from the internal delay, that amount of frequency range cannot be used on the power spectrum. Therefore, the reduction of the maximum detectable range occurs. These problems are shown by figures in Section VI.

### B. Concept of Proposed Down-conversion Technique

The proposed down-conversion technique is applied after the deramping process. When the last down-conversion is conducted with the last LO, the proposed technique uses the exact IF beat frequency of the leakage, $f_{IF\,beat,\,leakage} = f_{IF\,carrier} + f_{beat,leakage}$, and the exact constant phase, $\theta_{IF\,leakage}$, whereas the common technique uses the known IF carrier frequency, $f_{IF\,carrier} = f_{TX} - f_{RX}$. In other words, we propose the following LO.

$$LO_{proposed}(t) = D\cos\big(2\pi\big(f_{IF\,carrier} + f_{beat,leakage}\big)t$$

$$+ \theta_{IF\,leakage} + \varphi_{LO}(t)\big). \quad (9)$$

If (4) is mixed with (9) and passes a low pass filter, then the final signal, $z(t)$, can be expressed as follows:

$$z(t) = \underbrace{Z_{leakage}\cos\big(\varphi_{leakage}(t)\big)}_{Leakage}$$

$$+ \underbrace{Z_{target}\cos\big(2\pi f_{beat,target}t + \theta'_{target} + \varphi_{target}(t)\big)}_{Target}, \quad (10)$$

where $\theta'_{target} = \theta_{IF\,target} - \theta_{leakage}$. If the same approximation in (8) is applied in (10), $z(t)$ can be approximated as follows:

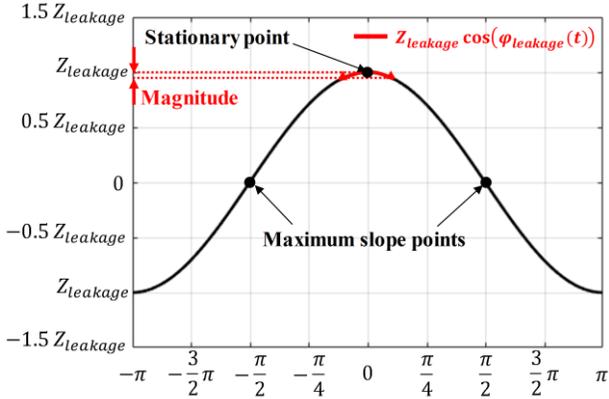

Fig. 3. Phase noise of the leakage on the stationary point.

$$z(t) \approx \underbrace{Z_{leakage}}_{Leakage} + \underbrace{Z_{target}\cos\left(2\pi f_{beat,target}t + \theta'_{target} + \varphi_{target}(t)\right)}_{Target}. \quad (11)$$

In (10), now that the leakage signal does not have any frequency and any constant phase, the term of the major cause of the phase noise skirt in (8) is removed. The only term associated with the phase noise of the leakage is the term which envelopes the phase noise of the leakage in the cosine function in (10). This term can be approximated to 1 for the same reason as in (8), then the leakage is represented as a DC value like as (11). Therefore, the proposed technique can resolve the problem of the phase noise of the leakage.

Even though we do not use the approximation, the effect of the proposed technique can be explained by Fig. 3. Through the proposed technique, the phase noise of the leakage is concentrated on the stationary point of the cosine function, because the leakage signal has no frequency and no constant phase. In case of the common technique, the beat signal, $y_{leakage}(t)$, has the beat frequency, so the phase noise of the leakage can tremble at every point on the cosine function, which includes the maximum slope points. On the other hand, in case of the proposed technique, the phase noise of the leakage trembles only at the stationary point. Therefore, the magnitude of the phase noise is significantly decreased, and the mitigation of the leakage signal is possible.

Also, the beat frequency corresponding to the internal delay is removed, so we can avoid the reduction of the maximum detectable range. When comparing the beat signal for the target from the common technique and the proposed technique, there is only difference in the constant phase. Hence, the power of the target is the same in both techniques.

The concept of the proposed technique has been explained in continuous time domain for the convenience of comparison with the common technique. However, in practice, we implement the proposed technique through DSP algorithm. Detailed explanations are provided in Section III.

## III. Implementation of Proposed Down-conversion Technique

One of the most important steps in the proposed down-conversion technique is finding the exact IF beat frequency, $f_{IF\ beat,\ leakage}$, and the exact IF constant phase, $\theta_{IF\ leakage}$, of the leakage. However, it is hard to anticipate these values. For example, a designer aims to make an analog LO whose frequency is 1 GHz, but actual frequency would be 999.99 MHz. Like this, the actual value of $f_{IF\ carrier}$ slightly differs from the value intended by the designer. Also, it is difficult to expect the exact value of $f_{beat,leakage}$ which depends on the internal delay. Therefore, the proposed technique carries these works in digital domain. IF beat signal, $x(t)$ in (4), is directly sampled by ADC. Fig. 4 shows the block diagram of the proposed technique.

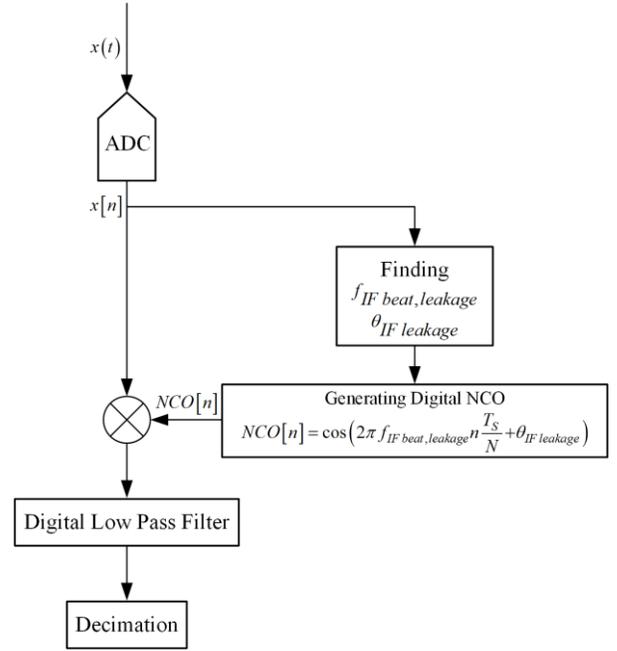

Fig. 4. Block diagram of the proposed down-conversion technique.

### A. Frequency Planning and Sampling

In the proposed technique, the sampling and the frequency planning for the IF beat signal should be conducted carefully. If the frequency planning and the sampling frequency is not considered cautiously, the desired frequency domain could be critically damaged when the multiplication for the last down-conversion is carried out. Generally, the sampling frequency is decided by the link budget analysis of FMCW radar. If the maximum physically detectable range is calculated by the link budget analysis, the beat frequency for this maximum range can also be calculated with the relation between the beat frequency and the distance. Then, twice the value of this beat frequency is going to be the sampling frequency by the Nyquist theorem. However, if the multiplication for the last down-conversion in the proposed concept is carried out with this sampling frequency, the desired frequency domain could be ruined. Fig. 5 shows this problem with an example. The IF beat signals are positioned carelessly on the power spectrum with unconsidered sampling frequency. When the multiplication process is conducted, undesired sum-terms are included in the desired frequency domain. Thus, targets around the sum-terms are hard to detect. Any digital filter cannot be used for the problematic part, because the target signals that were originally there are filtered together.

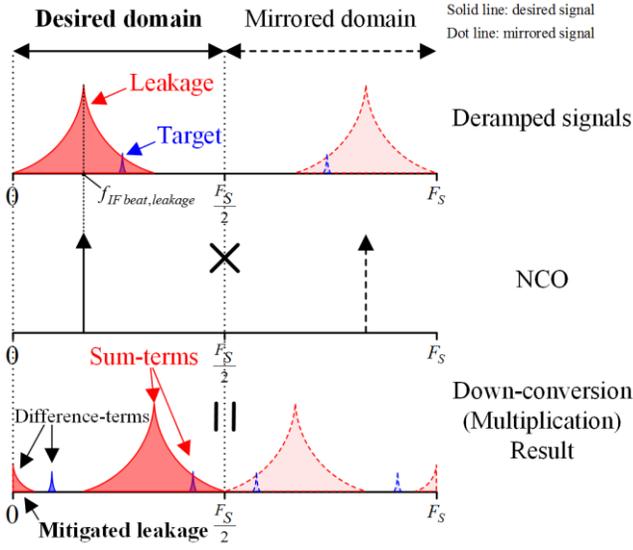

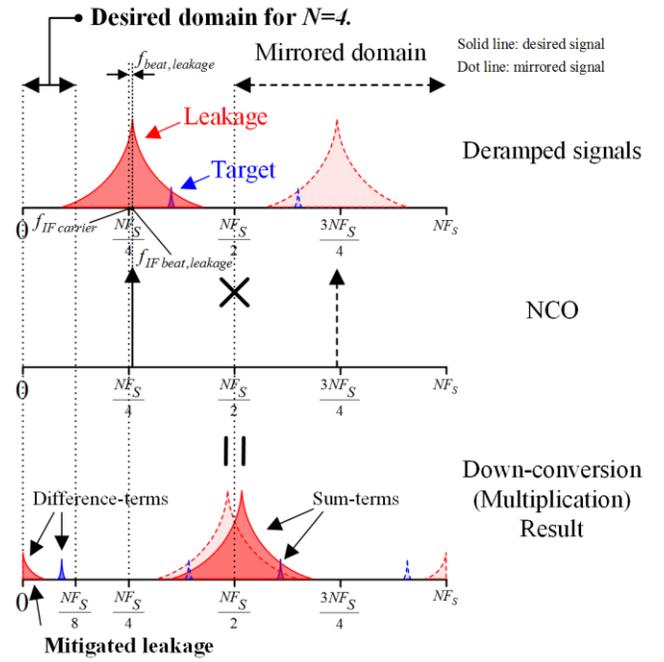

Fig. 5. Results of the last down-conversion in the proposed concept when IF beat signals are positioned carelessly on the power spectrum with unconsidered sampling frequency. Sum-terms which are results of multiplication (or mixing) process are included in the desired domain. Therefore, if targets are in this damaged part of the desired domain, they are hard to be detected.

Fig. 6. Results of the last down-conversion in the proposed concept when the proposed frequency planning and sampling is applied.

We propose a frequency planning and sampling scheme that can prevent the problem. Fig. 6 shows the scheme that we propose. We introduce an oversampling and suggest that the IF beat frequency, $f_{IF\ beat,leakage}$, be placed around a quarter of the oversampled frequency domain. In this way, we can keep the desired domain away from the sum-terms. Also, the sum-terms are located around the center of the digital IF frequency domain, which means that the sum-terms are now can be certainly removed by a low pass filter (LPF).

As the oversampling factor, $N$, becomes larger, the desired domain can be further away from the sum-terms and it reduces the threat that the sum-terms are to be aliased. However, if $N$ is too large, the cost of ADC increases for the high sampling rate. Small factor also gives a trouble. For example, if $N$ is 2, it can encounter the aliasing of the sum-terms. Therefore, we recommend the factor of 4 for $N$. In most case, the factor of 4 is enough to perform the proposed down-conversion technique.

It is difficult to anticipate the $f_{beat,leakage}$ before manufacturing an FMCW radar. However, $f_{IF\ beat,\ leakage}$ is mostly occupied by $f_{IF\ carrier}$, because $f_{IF\ carrier}$ is usually much larger than $f_{beat,leakage}$. Therefore, frequency planning that locates $f_{IF\ carrier}$ instead of $f_{IF\ beat,\ leakage}$ on a quarter of the oversampled frequency domain is reasonable. This explanation is reflected in Fig. 6. $f_{IF\ carrier}$ is $NF_S/4$, and the IF beat signal of the leakage whose frequency is $f_{IF\ beat,\ leakage}$ slightly misses $NF_S/4$ point. Nevertheless, the sum-terms are around the center of the oversampled frequency domain, so LPF can remove these obviously. Even in special cases that long internal delay, large $f_{beat,leakage}$, is inevitable [18], [19], the proposed technique also can be applied well by increasing $N$ a bit.

If we consider the undersampling which is also called as the bandpass sampling, the frequency planning can be realized on other locations, $NF_S(4M+1)/4$, where $M$ is natural number.

### B. Finding IF Beat Frequency and IF Constant Phase of Leakage

The very large power of the leakage signal is an obvious problem. However, we change this problem to a solution for finding the $f_{IF\ beat,\ leakage}$ and the $\theta_{IF\ leakage}$. We use the fact that the magnitude of the leakage signal is much larger than that of target signal [10], [11]. After applying the fast Fourier transform (FFT) at the oversampled IF beat signal, $x[n]$, we transform the FFT result to the power spectrum and the phase response. When the FFT is applied, we use the zero-padding to minimize an error caused by insufficiently spacing in the frequency domain. The FFT with zero-padding approaches to the result of discrete time Fourier transform, so it helps find the real location of the signal peak [21], [22]. Then, on the power spectrum, we find the peak which represents the maximum power around a quarter of the oversampled frequency domain. The index number, $k_{IF\ leakage}$, from the maximum peak can be extracted. Finally, we can find out the $f_{IF\ beat,\ leakage}$ and $\theta_{IF\ leakage}$ as follows:

$$k_{IF\ leakage} = \underset{\frac{NF_S}{4}<k<\frac{NF_S}{2}}{\arg\max}\left|X[k]\right|^2,$$

$$f_{IF\ beat,leakage} = \frac{NF_S}{NFFT}\left[k_{IF\ leakage}-1\right],$$

$$\theta_{IF\ beat,leakage} = \angle X\left[k_{IF\ leakage}\right], \quad (12)$$

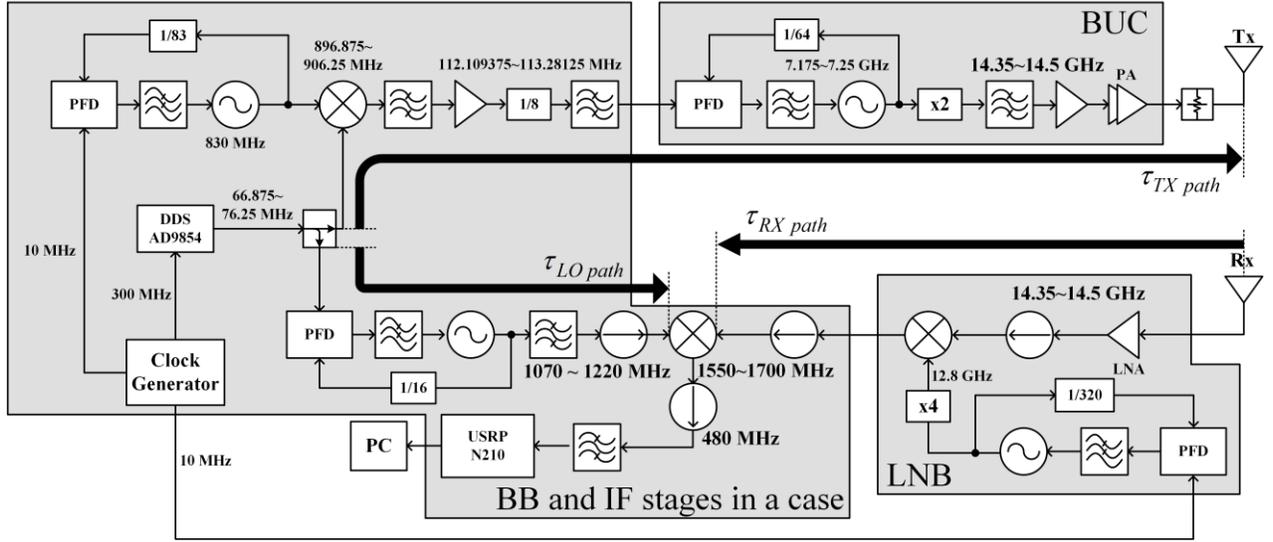

Fig. 7. Block diagram of the *Ku*-band FMCW radar system.

where $X[k]$ is the result of $NFFT$-point FFT of $x[n]$, $NFFT$ is the total number of real samples and zero-pads, $NF_S$ is the sampling frequency for the oversampling, and $\angle X$ is the phase response of $X[k]$.

### C. Down-conversion with Digital NCO, Filtering and Decimation

The oversampled IF beat signal, $x[n]$, can be written as follows:

$$x[n] = \frac{AB}{2}\cos\left(2\pi f_{IF\,beat,leakage}\, n\frac{T_S}{N} + \theta_{IF\,leakage}\right.$$
$$\left. + \varphi_{IF\,leakage}\left(n\frac{T_S}{N}\right)\right)$$
$$+ \frac{AC}{2}\cos\left(2\pi\left(f_{IF\,beat,leakage} + f_{beat,target}\right)n\frac{T_S}{N} + \theta_{IF\,target}\right.$$
$$\left. + \varphi_{IF\,target}\left(n\frac{T_S}{N}\right)\right), \quad (13)$$

where $T_S = 1/F_S$ is the sampling interval.

With the extracted $f_{IF\,beat,\,leakage}$ and $\theta_{IF\,leakage}$, the following digital NCO can be generated.

$$NCO[n] = \cos\left(2\pi f_{IF\,beat,leakage}\, n\frac{T_S}{N} + \theta_{IF\,leakage}\right) \quad (14)$$

Then, the last down-conversion can be conducted by multiplying $x[n]$ and $NCO[n]$. A digital LPF is used to get rid of the sum-terms and to pass the difference-terms in the desired domain. After the multiplication and the filtering, the signal can be expressed as follows:

$$\psi[n] = \frac{AB}{4}\cos\left(\varphi_{IF\,leakage}\left(n\frac{T_S}{N}\right)\right)$$

### TABLE I
### SPECIFICATIONS OF THE *Ku*-BAND FMCW RADAR

| Property | Value | Units |
|---|---|---|
| Operating frequency | 14.35-14.50 | GHz |
| Transmit power | 20 | dBm |
| Antenna gain | 16 | dBi |
| Sweep bandwidth | 150 | MHz |
| Range resolution | 1 | m |

$$+ \frac{AC}{4}\cos\left(2\pi f_{beat,target}\, n\frac{T_S}{N} + \theta'_{target} + \varphi_{IF\,target}\left(n\frac{T_S}{N}\right)\right) \quad (15)$$

Now that the major role of the oversampling is done, decimation by the factor of $N$ is carried out as the last work to avoid needless waste of the memory. The final output signal can be written as follows:

$$\zeta[n] = \underbrace{\frac{AB}{4}\cos\left(\varphi_{leakage}[n]\right)}_{Leakage}$$
$$+ \underbrace{\frac{AC}{4}\cos\left(2\pi f_{beat,target}\, n + \theta'_{target} + \varphi_{target}[n]\right)}_{Target}, \quad (16)$$

where $\varphi_{leakage}[n] = \varphi_{leakage}(nT_S)$, $\theta'_{target} = \theta_{IF\,target} - \theta_{leakage}$, and $\varphi_{target}[n] = \varphi_{target}(nT_S)$. The beat signal of the leakage does not have any frequency and any constant phase. Therefore, the term of the major cause of the phase noise skirt is removed, and the phase noise of the leakage is concentrated at the stationary point of the cosine function. This makes the phase noise of the leakage significantly mitigated. The internal delay is compensated by removing the beat frequency of it, so the reduction of the maximum detectable range can be prevented. Because only the constant phase changes in the beat signal of the target, its power remains the same.

## IV. RADAR SYSTEM

In order to verify the performance of the proposed down-conversion technique through experiments, a *Ku*-band FMCW radar was used. The *Ku*-band FMCW radar we used has been continuously upgraded for the drone detection in Microwave & Antenna laboratory at Korea Advanced Institute of Science and Technology (KAIST). The specifications of the *Ku*-band FMCW radar are listed on Table I.

Fig. 7 shows the block diagram of the *Ku*-band radar. The radar mainly consists of baseband (BB) & IF stages, block up converter (BUC), low-noise block (LNB), and antennas. The BB and IF stages are placed together in a metal case. A direct digital synthesizer (DDS), Analog Device AD9854 is used to generate the LFM signal with high linearity. An attenuator is added between the BUC and the TX antenna to adjust the transmit power and mitigate the reflected waves due to the mismatches between the PA and the Tx antenna. For the TX and RX antennas, corrugated conical horn antennas were used. High-efficiency and rotationally symmetric beam pattern can be obtained with the corrugated conical horn antenna [23]. The *Ku*-band radar includes a software defined radio (SDR), Ettus USRP N210. In the experiments, we used USRP N210 for the positioning of $f_{IF\ carrier}$ and the oversampling. The remaining procedures of the proposed technique after the oversampling were conducted through MATLAB in a mini-PC. For the digital LPF, we implemented the zero-phase digital filtering with the infinite-duration impulse response (IIR) filter. In terms of computation time and memory requirements, the IIR filter is usually more efficient than the finite-duration impulse response (FIR) filter [24]. Also, the phase distortion problem of the IIR filter can be alleviated by the zero-phase digital filtering, so the zero-phase digital filtering with the IIR filter is useful for the real-time processing [25].

## V. EXPERIMENTS

Radar set-up for the experiments is shown in Fig. 8. The quasi-monostatic type was selected by considering both the monostatic and the bistatic types of the radar. The distance between two antennas was 25 cm. The antennas were set up to look at the sky and experiments were carried out on the rooftop of a building in KAIST. Therefore, we created a situation that the received signals are only the leakage and the reflected signal by a target we set. For the target, we used a mini-drone, DJI Spark. There were two experiments. *Experiment A* is to verify that the proposed technique can mitigate the leakage by attenuating the phase noise of it. *Experiment B* is to demonstrate that the reduction of the maximum detectable range can be prevented by compensating the internal delay with the proposed technique. The radar parameters of each experiment are listed on Table II. The specifications of the digital LPF are shown in Table III.

### A. Experiment A: Mitigation of Leakage Signal

In this experiment, we focused on how much the proposed technique can reduce the phase noise of the leakage, namely how much the noise floor can be reduced. We created two situations for *Experiment A*. First, we operated the radar without any target, so that only the leakage would exist. In this condition, by comparing the power spectra from the common technique and from the proposed technique, we can check the reduced noise floor clearly. Second, we intentionally placed the mini-drone, Spark, in diagonal direction to make Spark out of the maximum beam axis. Fig. 9-(a) shows the second situation. Location information is from GPS in Spark. In this way, the target signal becomes weak, and it can be buried under the phase noise skirt. If the proposed technique works well, the target signal would not be buried under the phase noise skirt anymore.

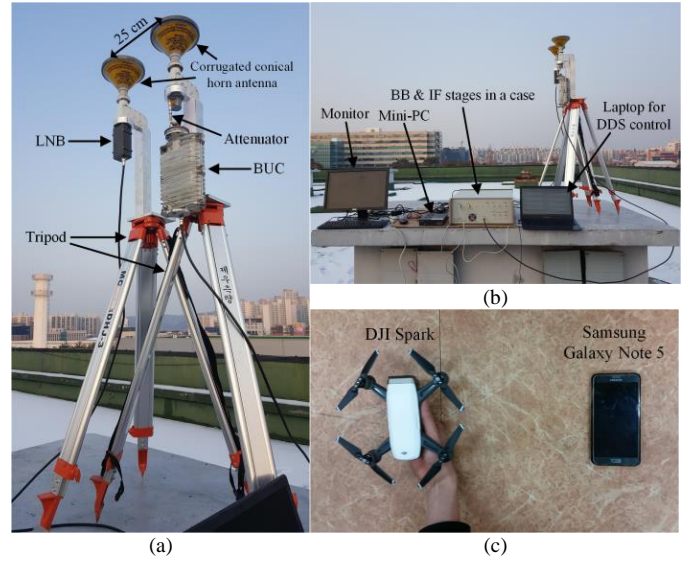

Fig. 8. Radar set-up and the target. (a) shows RF front-end and (b) shows the other parts of the radar. The target, DJI Spark is shown in (c). Spark can be held in the palm of one hand, and its size is similar with the smart phone.

TABLE II
FMCW RADAR PARAMETERS ACCORDING TO EXPERIMENT

| Parameters | Experiment A | Experiment B |
|---|---|---|
| Sweep period ($T$) | 860 us | 300 us |
| Frequency of LO in USRP N210 | 482.5 MHz | 480.5 MHz |
| The IF carrier frequency ($f_{IF\ carrier}$) | 2.5 MHz | 0.5 MHz |
| Oversampling frequency ($NF_s$) | 10 MHz | 2 MHz |
| Oversampling/Decimation factor ($N$) | 4 | 4 |
| Sampling frequency | 2.5 MHz | 0.5 MHz |
| # of samples in a chirp | 2150 | 150 |
| # of meaningful samples in a chirp | 2048 | 128 |
| Maximum detectable range | 1075 m | 75 m |
| Apparent range resolution | 1.0498 m | 1.1719 m |

TABLE III
SPECIFICATIONS OF DIGITAL LPF

| Specifications | Experiment A | Experiment B |
|---|---|---|
| Type of digital filter | Butterworth, IIR | Butterworth, IIR |
| Order | 11th | 11th |
| Stability | Stable | Stable |
| Sampling frequency | 10 MHz | 2 MHz |
| Passband frequency | 1.875 MHz | 0.375 MHz |
| Stopband frequency | 2.5 MHz | 0.5 MHz |
| Passband attenuation | 1 dB | 1 dB |
| Stopband attenuation | 30 dB | 30 dB |

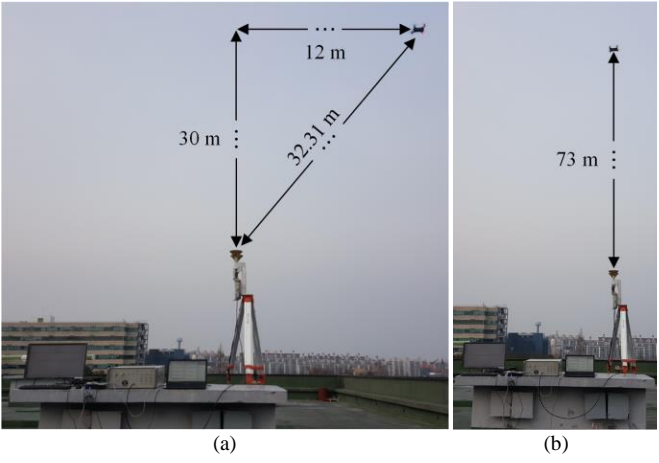

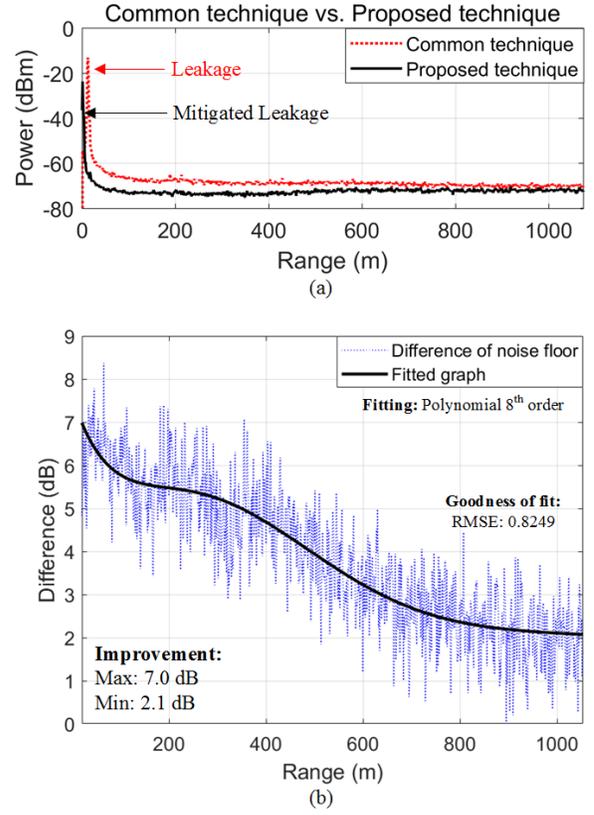

Fig. 9. Experiment scenarios. (a) shows the second situation of *Experiment A* and (b) shows *Experiment B*.

### B. Experiment B: Compensation of Internal Delay

If the maximum detectable range is reduced by the internal delay in the common technique, the target at a distance that should have been detected may not be detected. On the other hand, the proposed technique will detect the target. In order to demonstrate this situation, we have to place the target near the maximum detectable range. However, it is difficult to place the target in far away in practice. Horizontally, there can be many obstacles between the target and the radar. Vertically, in case of drones of DJI, the maximum flight altitude is limited for safety. Also, the maximum flight altitude is restricted by a flight safety law in South Korea. Therefore, we changed the radar parameters and specifications to make the short-range radar. This can be checked in Table II and Table III. The theoretical maximum detectable range is 75 m, and we placed Spark at the height of 73 m above the antennas. Fig. 9-(b) shows this situation.

## VI. RESULTS AND DISCUSSION

### A. Results and Discussion on Experiment A

The results of the first situation are shown in Fig. 10. Because we did not put any target in the first situation, there is only the leakage signal on the power spectrum. For the clear comparison between the results of the common technique and proposed technique, we took an average on the power spectra from 100 chirps to reduce variance of the noise floor. Fig. 10-(a) shows that the proposed technique mitigates the leakage signal. The phase noise skirt is remarkably decreased, and the entire noise floor is lowered. Not only the noise floor around the near-distance but also the noise floor around the far-distance are reduced. We calculated the difference of the noise floor and applied the curve fitting to estimate the degree of improvement. For the fitting, the 8th order polynomial function was used. As shown in Fig. 10-(b), the maximum degree of improvement is about 7.0 dB in the near-distance region, and the minimum degree of improvement is about 2.1 dB in the far-distance region.

Fig. 11 shows the results of the second situation. On the power spectrum through the common technique, i.e., Fig.

11-(a), the target signal is buried under the phase noise skirt and there is only the leakage signal. On the other hand, on the power spectrum from the proposed technique, i.e., Fig. 11-(b), the phase noise skirt is significantly decreased, and the target signal certainly exists. The comparison of the power spectra is represented in Fig. 11-(c).

In the power spectra from the proposed technique in Fig. 10 and Fig. 11, there is a part of the phase noise skirt that is not completely mitigated and remains a bit. Of course, this can happen, because the approximated result, (11), is the approximated one as the word itself. It is practical to consider (10) and Fig. 3, when we take care of the practical environment. Meanwhile, the leakage paths can be the one of the reasons for this phenomenon. There is not the only one spatial path for the leakage, but there are also many other paths in practical environment. The delay is slightly different for each path, so there is not just one beat frequency for the leakage. Even so, the proposed technique focuses the leakage signal whose magnitude is biggest. Therefore, it is true that the most dominant leakage signal from the most dominant leakage path can certainly be mitigated by the proposed down-conversion technique.

### B. Results and Discussion on Experiment B

To compare the results clearly, an average of the power spectra from 100 chirps was taken for Fig. 12. In Fig. 12-(a), it can be said that the power spectrum from the proposed technique normally detects Spark, when we consider the true distance of Spark is 73 m. There can be small error in the measured distance because of the apparent range resolution, the GPS error, and the distance between two antennas. However, the measured distance of the target on the power spectrum from the common technique is too different from the true distance. Fig. 12-(b) explains why this occurs. The common technique does not care about the internal delay which leads to the additional beat frequency. Of course, this additional beat frequency is added to the beat frequency of the target, then the target exceeds the half of the sampling frequency. Therefore, it can be said that the maximum detectable range is reduced by the amount of the additional beat frequency. If the distance between the antennas are close enough, the additional beat frequency is almost same as the beat frequency of the leakage. Because the beat frequency of the leakage is represented as 12.89 m in the range domain, the real target signal on the power spectrum from the common technique should be around 12.89 m + 73 m = 85.89 m. Therefore, the signal peak whose measured distance is 85.55 m in the mirrored domain is the real target signal. However, it exceeds the half of the sampling frequency which is represented as 75 m in range domain. Thus, its aliased signal is created on 64.45 m = 75 m − (85.55 m − 75 m). The target signal is aliased if the distance of the target is over 62.11 m = 75 m − 12.89 m. Therefore, in fact, it can be said that the actual maximum detectable range is reduced from 75 m to 62.11 m in the common technique. On the other hand, because the internal delay is compensated through the proposed technique, it is shown that the proposed technique prevents the reduction of the maximum detectable range in Fig 12-(a) and Fig 12-(c). Also, since the proposed technique changes only the constant phase in the target signal, there is little change in the power of the target signal. This is verified in Fig 12-(a).

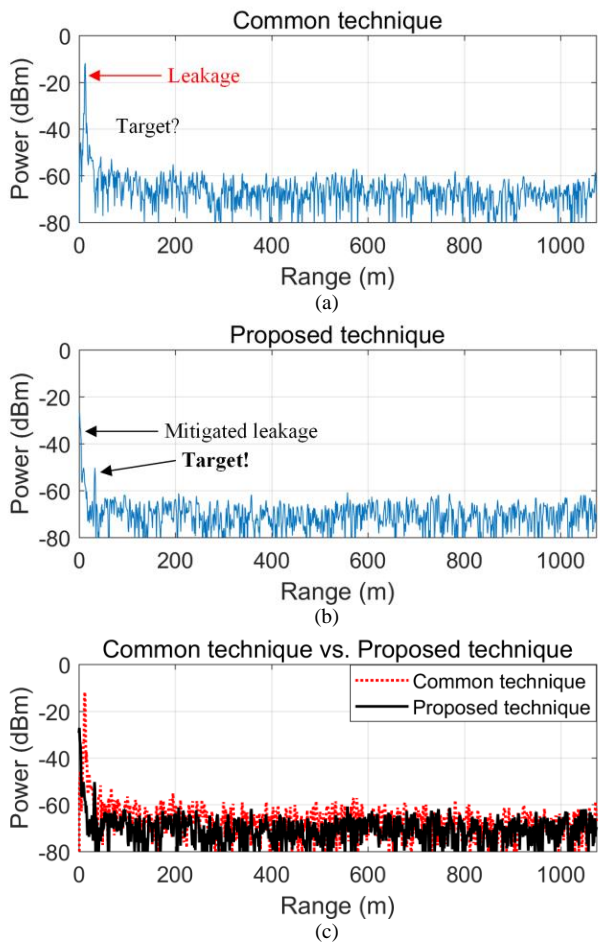

Fig. 11. Results of the second situation in *Experiment A*. (a) and (b) are the power spectra from the common technique and from the proposed technique respectively. (c) compares these power spectra.

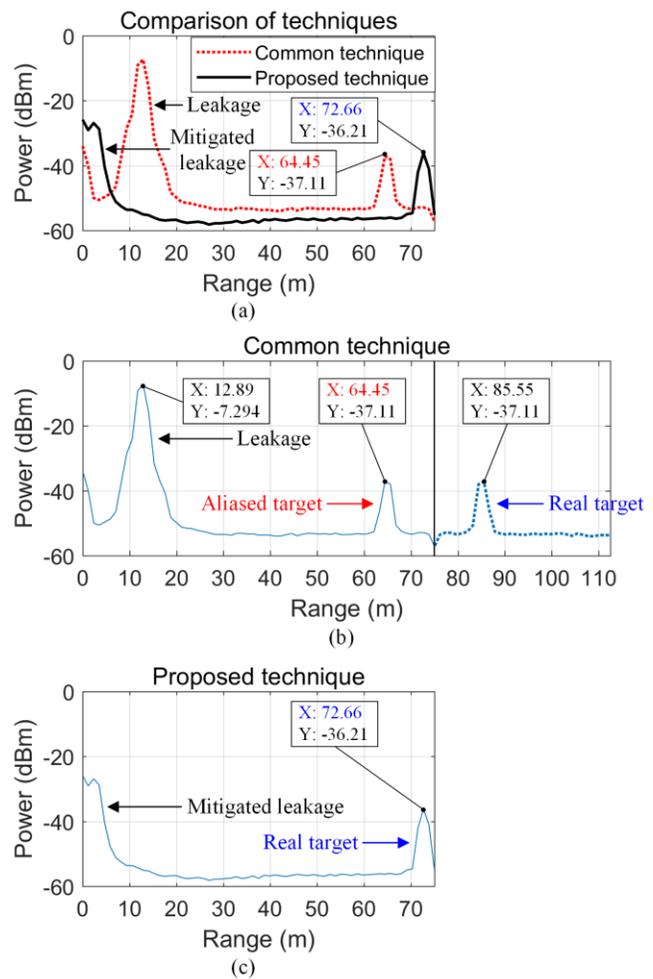

Fig. 12. Results of *Experiment B*. (a) compares the power spectra from the common technique and the proposed technique. (b) and (c) shows the power spectrum from the common technique and the proposed technique respectively. (b) helps to explain the reason why the common technique reduces the maximum detectable range.

## VII. CONCLUSION

A new down-conversion technique to mitigate the leakage and compensate the internal delay in FMCW radar has been explained in detail and verified with the experiments. The proposed technique can be implemented through the frequency planning and DSP without additional parts. Thanks to the proposed technique, the phase noise skirt is significantly reduced, so the noise floor decreases. The maximum degree of improvement for the noise floor is about 7.0 dB in the near-distance region, and the minimum degree of improvement is about 2.1 dB in the far-distance region. Also, the proposed technique can prevent the reduction of the maximum detectable range by compensating the internal delay. In the target signal, only the constant phase is changed through the proposed technique. Therefore, the aforementioned benefits of the proposed technique can be obtained without affecting the target signals on the power spectrum.